# The Specialized High-Performance Network on Anton 3

Keun Sup Shim, Brian Greskamp, Brian Towles,[1] Bruce Edwards, J.P. Grossman,[1] David E. Shaw[2]

D. E. Shaw Research, New York, NY 10036, USA.

{Keun.Sup.Shim, Brian.Greskamp, David.Shaw}@DEShawResearch.com

*Abstract*—Molecular dynamics (MD) simulation, a computationally intensive method that provides invaluable insights into the behavior of biomolecules, typically requires large-scale parallelization. Implementation of fast parallel MD simulation demands both high bandwidth and low latency for inter-node communication, but in current semiconductor technology, neither of these properties is scaling as quickly as intra-node computational capacity. This disparity in scaling necessitates architectural innovations to maximize the utilization of computational units. For Anton 3, the latest in a family of highly successful special-purpose supercomputers designed for MD simulations, we thus designed and built a completely new specialized network as part of our ASIC. Tightly integrating this network with specialized computation pipelines enables Anton 3 to perform simulations orders of magnitude faster than any general-purpose supercomputer, and to outperform its predecessor, Anton 2 (the state of the art prior to Anton 3), by an order of magnitude. In this paper, we present the three key features of the network that contribute to the high performance of Anton 3. First, through architectural optimizations, the network achieves very low end-to-end inter-node communication latency for fine-grained messages, allowing for better overlap of computation and communication. Second, novel application-specific compression techniques reduce the size of most messages sent between nodes, thereby increasing effective inter-node bandwidth. Lastly, a new hardware synchronization primitive, called a *network fence*, supports fast fine-grained synchronization tailored to the data flow within a parallel MD application. These application-driven specializations to the network are critical for Anton 3's MD simulation performance advantage over all other machines.

## I. INTRODUCTION

With molecular dynamics (MD) simulations, scientists can study the behavior of biological molecules (e.g., proteins, lipids, and nucleic acids) by computationally predicting their motion at the atomic scale. Although these simulations have provided tremendous value in both basic science and drug discovery, they require an enormous amount of computation, and performing simulations at the scale and speed necessary to address relevant research questions within practical timelines remains a highly challenging problem [1][2].

The Anton supercomputers are a series of special-purpose machines designed to accelerate MD simulations. When Anton 1 and Anton 2 were deployed, each exceeded the performance of the fastest general-purpose machine of its time by at least two orders of magnitude [3]–[5]. The latest generation in the Anton family, Anton 3, achieves order-of-magnitude improvements in time-to-solution over Anton 2, and outperforms the current state-of-the-art general-purpose supercomputers by over 100-fold across a wide range of chemical system sizes [6][7].

Table I shows that compared to the Anton 2 ASIC, the Anton 3 ASIC delivers roughly 24 times more maximum throughput for the most computationally expensive task in MD—computing the forces between two interacting atoms (we also refer to this as computing pairwise interactions). Translating this raw intra-node computational throughput into better overall simulation performance requires a high-performance network that can utilize the compute resources efficiently. Scaling the inter-node bandwidth at low latency, however, is a challenge because packaging constraints limit the number of high-speed off-chip lanes on the chip, and per-lane bandwidth is growing slowly for low-latency SERDES due to the physical limitations of long-distance electrical signaling (as evidenced by the "Number of SERDES" and "SERDES Per-Lane Bandwidth" in Table I). Compared to the 24-fold improvement in raw compute bandwidth from Anton 2 to Anton 3, the 2.1-fold improvement in the total inter-node bandwidth motivates the need for network specialization to better utilize the off-chip bandwidth. Furthermore, the network must enable efficient synchronization among the large number of compute units in Anton 3 machines in order to effectively coordinate parallel computation.

Communication challenges are especially acute in MD simulation because: (1) parallel implementations of MD require significant inter-node communication, as atom positions and computed forces have to be frequently exchanged between nodes; and (2) some synchronizations are essential during the MD time steps, and these synchronizations contribute to the critical path if not well optimized. Designing a high-performance network is thus critical for effective parallelization of MD. To this end, we designed and implemented the Anton 3 network with the following features:

1) *Fast end-to-end inter-node communication*: The Anton 3 network is designed to minimize inter-node communication latency for fine-grained messages. The end-to-end latency between cores can be as low as 55 ns for neighboring nodes (considerably lower than other high-performance computing networks), with an off-chip per-hop latency of 34.2 ns. Fast inter-node communication for fine-grained messages is key for optimizing MD performance, as it allows for better overlap of computation and communication.

2) *Application-specific compression*: Atom positions in simulations change slowly and smoothly over time, offering an opportunity for data compression. We thus designed and built the *particle cache*, a lossless compression scheme at off-chip boundaries. By caching atom positions at both ends of an I/O channel, it is only necessary to transmit the position *differences* for each time step, thereby saving communication bandwidth. Combining the particle cache with *Interleaved Non-Zero encoding* (*INZ*), a compression scheme we developed for payloads with small absolute values, the inter-node communication traffic can be reduced by twofold, mitigating the off-chip bandwidth scaling problem.

3) *Fast fine-grained synchronization*: We have implemented a novel hardware synchronization primitive within the network, which we call a *network fence*, that supports fast synchronization among compute units while requiring little area on the chip.

TABLE I. KEY FEATURES FOR THE THREE ANTON ASICs.

|  | Anton 1 | Anton 2 | Anton 3 |
|---|---|---|---|
| Power-on Year | 2008 | 2013 | 2020 |
| Process Technology (nm) | 90 | 40 | 7 |
| Die Size (mm$^2$) | 305 | 408 | 451 |
| Clock Rate (GHz) | 0.970 | 1.65 | 2.80 |
| Maximum Pairwise Interaction Throughput (GOPS) | 31 | 251 | 5914 |
| Number of SERDES | 66 | 96 | 96 |
| SERDES Per-Lane Bandwidth (Gb/s) | 4.6 | 14 | 29 |
| Inter-node Bidir Bandwidth (GB/s) | 76 | 336 | 696 |

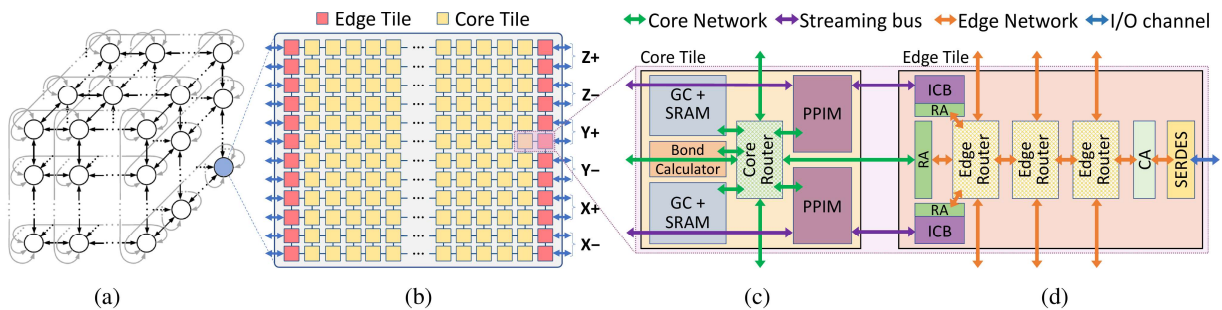

Figure 1. (a) Anton 3 machines comprise up to 512 ASICs connected in a 3D torus network. (b) The ASIC uses a tiled layout containing a 24 × 12 array of Core Tiles, with 12 Edge Tiles along each of the left and right edges. X+, X−, Y+, Y−, Z+, and Z− are the six neighbors in the torus. (c) Block diagrams of the Core Tile and (d) Edge Tile.

## II. BACKGROUND

### A. Molecular Dynamics Simulation

MD simulation models the motion of atoms in a chemical system using a series of discrete time steps. A time step consists of first calculating the forces exerted on each atom by other atoms in the system, and then updating the velocities and positions of all atoms according to the classical laws of motion. MD simulations often require several billion sequential time steps to reach the timescales at which many scientifically interesting phenomena start to occur.

One of the most computationally intensive tasks in MD simulation is the computation of range-limited pairwise interactions—the forces between all pairs of atoms separated by less than an established cutoff radius—at every time step. The MD computation can be parallelized across multiple nodes by spatially partitioning the chemical system into boxes, and assigning each box to a node (its *Home Node*) that is responsible for updating the positions of the atoms within the assigned box (the *Home Box*). Because the computation of range-limited pairwise interactions requires the positions of atoms not only within the Home Box, but also within neighboring boxes (i.e., on remote nodes), substantial inter-node communication is necessary. This communication requires significant off-chip bandwidth, ideally with very low latency.

### B. Anton 3 Architecture Overview

Anton 3 machines comprise up to 512 nodes (with a single ASIC per node) that are physically connected by integrated high-speed network links to form a 3D torus topology (Figure 1a). Each ASIC (Figure 1b) contains two types of tiles: *Core Tiles* (Figure 1c) and *Edge Tiles* (Figure 1d). The Core Tiles, organized as an array of 12 rows and 24 columns, perform the MD computations. 24 Edge Tiles flank the left and right sides of the Core Tile array (i.e., 12 on each side), where they primarily manage inter-node communication. The chip contains 96 bi-directional SERDES lanes distributed evenly among the Edge Tiles, each operating at 29 Gb/s per direction. These lanes connect each ASIC to its six neighbors (with 16 SERDES per neighbor) in the 3D torus network, providing a total bandwidth of 5.6 Tb/s.

The Core Tile contains: (1) two MD-optimized general-purpose processors called *Geometry Cores* (GCs), each paired with a globally addressable on-chip memory block with 128 KB of SRAM; (2) two *Pairwise Point Interaction Modules* (PPIMs), which include several specialized arithmetic pipelines responsible for range-limited pairwise interactions; (3) the *Bond Calculator* (BC), which computes forces between pairs of atoms bonded directly, or through one or two intervening atoms; and (4) the *Core Router*, which connects the components within a Core Tile, and also connects the Core Routers in neighboring tiles to form a 2D mesh on-chip network called the *Core Network*. In addition, dedicated streaming buses run horizontally across the chip through PPIMs, carrying atom positions and forces for the evaluation of pairwise interactions.

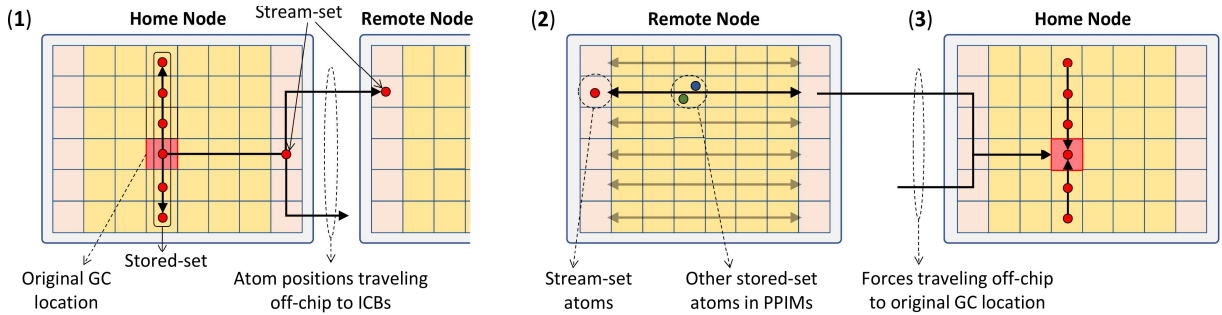

Figure 2. The three main steps in the data flow for computing range-limited pairwise interactions on Anton 3. (The numbers of rows and columns for Core Tiles (yellow) and Edge Tiles (pink) have been reduced for ease of presentation.) (1) An atom position is multicast from a GC to PPIMs within its column (as a stored-set atom) and to ICBs on all nodes on which that atom might have an interaction (as a stream-set atom). (2) ICBs stream an atom position across a row of PPIMs, where pairwise forces are calculated with stored-set atoms. (3) Stream-set and stored-set forces for an atom then return to the original GC, where all of the forces are summed.

Each Edge Tile contains three *Edge Routers*; the connections between vertically adjacent Edge Tiles thus form a mesh network of 12 rows and 3 columns on both the left and right edges of an ASIC (we refer to these as *Edge Networks*). The Edge Networks interface with the SERDES channels through *Channel Adapters* (CAs), and with the Core Network through *Row Adapters* (RAs). The primary function of the Edge Networks is to implement routing for the inter-node 3D torus network. Each Edge Tile also contains two *Interaction Control Blocks* (ICBs) that receive packets from their Edge Network, buffer the packets within local ICB memory, and send them across their row's streaming buses to participate in the calculation of pairwise interactions in the PPIMs. Each ICB has a Row Adapter to connect to the Edge Network.

### C. Data flow for Calculating Pairwise Interactions on Anton 3

The implementation of parallel MD simulation on Anton 3 guarantees that pairwise interactions between two atoms are computed on a node that contains either one or both of the interacting atoms within its Home Box. Below, we describe the data flow for calculating pairwise interactions of atoms on Anton 3, with a focus on interactions between atoms that reside on different nodes.

At the beginning of a time step, each GC holds information regarding a different subset of the atoms within a given Home Box. From this point, there are three main steps of data flow (as illustrated in Figure 2):

1) Each GC broadcasts its atom positions to PPIMs within its column, where they are held as *stored-set atoms*. The atom positions are also sent to the Edge Tiles as *stream-set atoms*, and from there they are multicast to ICBs on all nodes on which those atoms might have an interaction.

2) As positions for these stream-set atoms arrive at the ICBs, they are streamed through a row of PPIMs using the streaming bus, and interact with stored-set atoms in PPIMs. These interactions result in forces being computed on both sets of atoms; the forces on the stored-set atoms are accumulated within the PPIM, while those on the stream-set atoms are streamed back along the streaming bus and returned to the GCs that originally sent the atom positions (possibly traveling off-chip).

3) Once all the forces for the stream-set atoms (i.e., *stream-set forces*) are returned, the accumulated forces for the stored-set atoms (i.e., *stored-set forces*) are unloaded from PPIMs and also sent back to their original GC locations for per-atom summation.[3]

GCs then perform integration to compute new velocities for all atoms based on the summed forces, use those velocities to update atom positions, and proceed to the next time step.

### III. FAST END-TO-END INTER-NODE COMMUNICATION

Reducing end-to-end inter-node communication latency is critical to achieving high performance for parallel MD simulations. End-to-end latency includes not only the raw hardware network latency, but also the time required for a source to initiate transmission of a message, and for the destination to finish receiving the message and be ready to perform computation on its contents. In this section, we detail the hardware elements in Anton 3 that help to minimize this latency.

### A. Counted Write and Blocking Read Synchronization

Counter-based, fine-grained synchronization is a key communication paradigm for the Anton ASICs [5][8]. The core principle is that the sender can send a remote memory-write message that increments an associated hardware counter at the receiver, and the counter can be used to detect that all data required has been received. In Anton 3, this functionality is implemented using an 8-bit counter associated with each of the *quads* (each quad comprises four 32-bit values) inside SRAM memory blocks. Counted remote write messages to SRAM update the quad data which atomically increments this per-quad counter. Software running on GCs may issue a *blocking read* to a specific quad address within its local SRAM, and the blocking read will stall until the quad's counter has reached the counter threshold specified by the read; from the GC's point of view, this operation is no different than a high-latency read.

---

[3] Although not discussed in this paper, Anton 3 implements in-network hardware support for stored-set and stream-set atom multicast, as well as in-network reduction for summing stored-set forces.

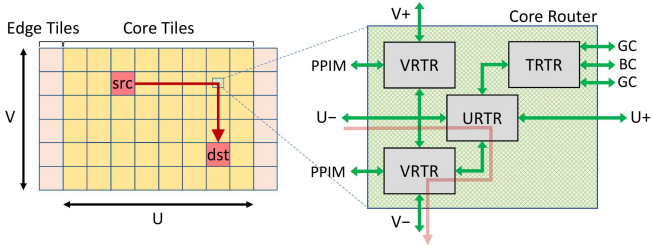

Figure 3. The Core Router within the Core Tile implements U→V dimension-order routing (as shown by the red arrow), and consists of four sub-routers (TRTR, URTR, and two VRTRs).

The purpose of blocking read synchronization is to minimize the arrival-to-use latency for data received over the network. In particular, it allows software handlers to start running before all of the input data has arrived. When a GC attempts to read data from SRAM that is not yet ready (i.e., the count is below the threshold), the read stalls and is only completed upon arrival of the data. One common use of counters during MD simulation is within the integration code, where each force for a given atom is accumulated into a quad in SRAM, which increments the corresponding counter. The integrator knows how many forces to expect for a given atom, and thus can use a blocking read to wait for the fully accumulated net force on the atom to be available.

### B. Packets and Routing

The Anton 3 network is designed to provide efficient support and low latency for fine-grained messages, using small, fixed-size network packets that comprise one or two flits (each flit contains 192 bits divided into a 64-bit header and a 128-bit payload). These small packets enable fast and efficient virtual cut-through flow control with small router input queues each holding eight flits per virtual channel (VC). In order to reduce per-hop latency on the critical timing path, all routers separate their control logic (which performs routing and arbitration for the next hop) from the packet datapath. This allows the packet data to lag its corresponding control information by two cycles.

*1) Core Network:* As shown in Figure 3, the 2D mesh Core Network handles intra-chip traffic using fixed U→V dimension-order routing with virtual cut-through flow control (we use U and V for the two mesh dimensions in order to distinguish from X, Y and Z of the inter-node 3D torus network dimensions). Packets targeting remote ASICs are routed directly to either edge of the chip, traveling along the U dimension only.

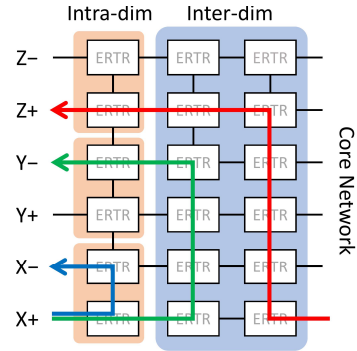

Figure 4. Examples of request packet routes within the Edge Network. Intra-dimensional traffic (e.g., the blue route) travels only within the outermost column of the Edge Routers (ERTRs), and the opposite directions of the same 3D-torus dimension (e.g., X+ and X−) are connected to adjacent rows. The remaining two columns are used by inter-dimensional traffic (e.g., the red and green routes).

In order to minimize latency, the Core Router adopts a partitioning approach similar to a dimension-sliced router [9], and is implemented using four sub-routers, each of which has at most four ports. Moreover, just two VCs suffice to avoid network deadlock between two classes of protocol traffic (requests and responses). There are three types of sub-routers—*URTR*, *VRTR*, and *TRTR*—that are microarchitecturally similar, but have distinct roles in the network. URTR and VRTR perform the inter-tile routing along the U and V dimensions, respectively. TRTR connects the GCs and BCs to the network and provides high bandwidth for local communication between those endpoints. As a whole, the Core Router has a per-hop latency of two cycles in the U direction, and five cycles in the V direction.

*2) Edge Network:* Inter-node routing in the 3D torus network of Anton 3 employs minimal, oblivious routing in which routes are randomized independent of network load, and packets follow a dimension-order route using any of the six possible dimension orders (i.e., XYZ, XZY, YXZ, YZX, ZXY, or ZYX). The routing is implemented within the Edge Network, which consists of three columns of Edge Routers, as previously described in Section II-B.

The routing policy for request packets is designed to achieve low latency through the Edge Network. Figure 4 illustrates examples of request packet routes within the Edge Network, which is physically partitioned by column between intra-dimensional traffic and inter-dimensional traffic. The outermost column of the Edge Routers is reserved for packets that are injected at the channels and traveling to another channel in the same torus dimension. This partitioning ensures that routing along a torus dimension requires minimal hops in the Edge Network. Other traffic, including packets injected from the Core Network and packets making a turn in the 3D torus, can use the remaining two columns in a randomized fashion for better load balance.

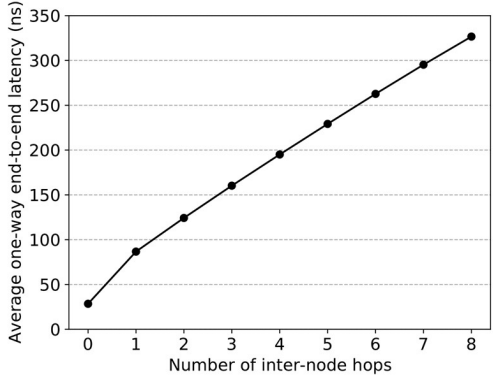

Figure 5. Average one-way end-to-end latency with 16 bytes of payload plotted against the number of inter-node hops, as measured on a 128-node Anton 3 machine.

To avoid network deadlock, the application-level protocol in Anton 3 requires separate traffic classes for requests and responses, for which torus routing would normally require four VCs for each class [10]. Instead, the number of VCs is reduced by introducing the restriction that all response packets must follow an XYZ dimension order, and by treating the torus network as a mesh network from the perspective of response packets. This enables deadlock avoidance using only a single VC for the response class, while having negligible impact on MD performance (as most traffic during simulations on Anton 3 is architected to belong to the request class). This amounts to a total of five VCs for the Edge Router, allowing the Edge Router to achieve a per-hop latency of three cycles.

*C. Evaluation*

End-to-end inter-node communication latency was measured by running a ping-pong test on a real ASIC. The test starts with software running on one GC (core A) sending a counted write of 16-byte data to memory associated with another GC (core B) on a remote ASIC. Software running on core B issues a blocking read to this local memory location, and upon receipt, it sends a counted write back to core A. Core A also has its blocking read issued for this message returned from core B. The ping-pong is complete when core A receives this packet and its blocking read is un-stalled, and the one-way end-to-end latency is computed as half the average time it takes to complete a single ping-pong.

Figure 5 plots the average one-way end-to-end latency of Anton 3 as a function of the number of inter-node hops. The latency was measured on a 128-node machine, and averaged over all possible GC pairs that are a given number of hops apart (because the location of cores within a chip affects intra-chip latency). The measurements were performed at the Anton 3 ASIC's typical operating clock frequency of 2.8 GHz with 29 Gbps channels, and show that a linear fit of 55.9 ns of fixed overhead plus 34.2 ns of per-hop latency is a good approximation of the average one-way end-to-end latency, except for the 0-hop case, which has lower latency because packets do not have to travel through the Edge Network and off-chip links.

The minimum inter-node latency measured for a single hop was approximately 55 ns, almost half that of the Anton 2 network (99 ns). Figure 6 shows how this latency is broken down among the endpoints and various network components. The improvement in the inter-node latency results from the tight integration between the network and the core (e.g., not using a communication library, such as MPI), which minimizes the sender and receiver overheads, and from network optimizations for fine-grained messages that reduce per-hop latency. Overall, the end-to-end inter-node latency of Anton 3 is substantially lower than typical inter-node message latencies on commodity networks, particularly between near-neighbor nodes (a dominant communication pattern in MD simulation). InfiniBand [11] and Intel Omni-path Architecture [12], for example, typically have around 1-μs one-way latency [13]–[15]. By way of comparison, another specialized custom network, Tofu interconnect D [16] (adopted in the Fugaku supercomputer [17][18]), has a minimum one-way latency of 490 ns, about nine times longer than the Anton 3 network.

IV. APPLICATION-SPECIFIC COMPRESSION

In this section, we describe interleaved non-zero encoding (INZ) and the particle cache, two compression techniques designed for the MD application to reduce off-chip network traffic.

*A. Interleaved Non-Zero Encoding*

In our MD simulations, flit payloads often contain three or four signed 32-bit word values (representing, for example, forces, charges, etc.). To reduce both electrical switching activity in the on-chip network and bandwidth over the I/O channels, the flit payload encoding is designed to optimize for the common case in which payloads have small absolute values. The encoding scheme reduces the number of bytes required to send data by maximizing the number of leading zeros in the payload to improve subsequent compression. Multiple compressed payloads and their accompanying headers are then densely packed (at byte granularity) into each fixed-length channel frame that traverses the off-chip interface. Utilization of off-chip bandwidth is thus improved by removal of the most significant zero bytes in each payload, allowing more data to fit into each channel frame.

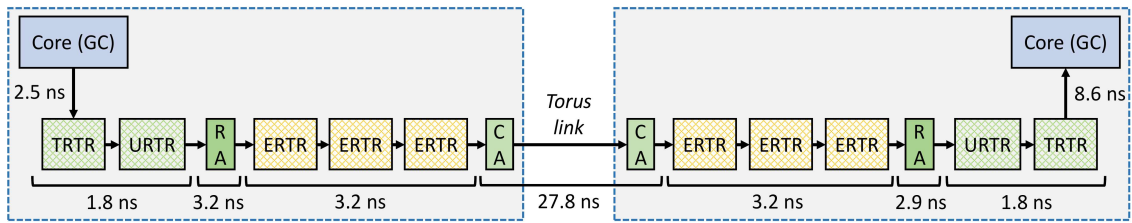

Figure 6. Detailed breakdown of the minimum inter-node, end-to-end latency of 55 ns. Abbreviations: ERTR, Edge Router; RA, Row Adapter; CA, Channel Adapter.

This encoding scheme, using an example with two 32-bit words, is illustrated in Figure 7. First, the most significant non-zero word is determined (there could be 0–4 non-zero words in our actual quad-word payloads). For each non-zero word $w$, the sign bit is moved to the least significant bit and the non-sign bits are conditionally inverted according to the sign bit, as in the following SystemVerilog function:

```
function logic [31:0] invert_word(logic [31:0] w);
    return {{31{w[31]}}^w[30:0],w[31]};
endfunction
```

Next, the non-zero words are interleaved bitwise (hence the scheme is referred to as interleaved non-zero encoding, or INZ). If there are no non-zero words, the total number of bytes in the payload is simply zero. Otherwise, the most significant non-zero word can be represented using two bits, and these are concatenated with the bit interleaved words. The number of non-zero bytes in the resulting vector are counted. If this vector is greater than 128 bits, the encoding is abandoned and the original data is used instead. In this special case, the number of valid bytes is set to 16. For the example in Figure 7, INZ-encoding of 8-byte data (two words) results in 5 bytes of leading zeros that can be eliminated during off-chip communication. In Anton 3, INZ-encoding or decoding of a 16-byte payload is done in a single cycle at 2.8 GHz.

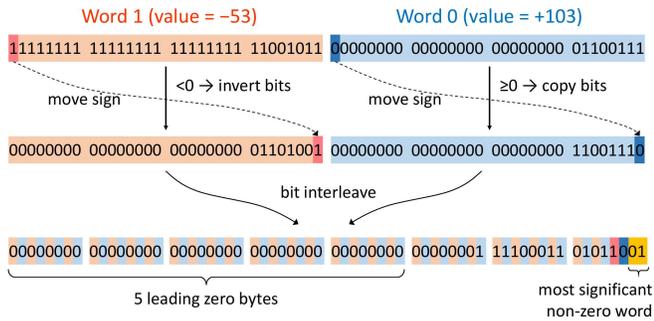

Figure 7. An example of INZ (interleaved non-zero encoding). The most significant byte is moved from byte 7 to byte 2 in this 8-byte data example.

### B. Particle Cache

As described in Section II-C, in order to calculate range-limited pairwise interactions during each MD time step, atom positions (also referred to as particle positions) need to be exported to neighbor nodes over I/O channels. The position export bandwidth is thus a critical factor for performance of parallel MD simulations, and the off-chip bandwidth should be utilized as efficiently as possible. To achieve this in Anton 3, we implemented the particle cache—a hardware mechanism that significantly reduces the amount of data required to communicate a particle position over off-chip channels. This technique was designed to take advantage of the fact that atom positions are exported over the same channels on multiple consecutive time steps, and although the positions themselves are not small values, they change slowly over time. In addition, there are a few fields in the position packet that remain static throughout the simulation. We can thus decrease off-chip communication by caching the atom information on the receive side, and sending only position deltas on most time steps. Because position deltas are generally small values, they exploit INZ compression more effectively.

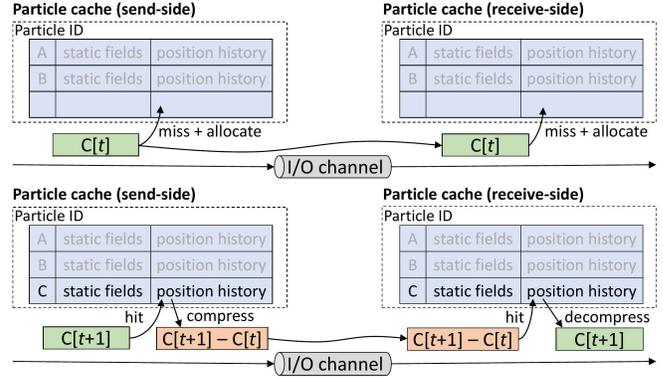

Figure 8. An example of particle cache operation. The top half of the diagram illustrates the case of a cache miss for C[$t$] (the position of particle C from time step $t$). The bottom half illustrates the case of a cache hit in the subsequent time step, with resulting compression and decompression.

*1) Implementation:* Figure 8 illustrates the high-level structure and basic operation of the particle cache. The particle cache is implemented within the Channel Adapter using two synchronized caches, with each sitting at either end of an I/O channel; the send-side cache is located before the I/O channel, and the receive-side cache is located after the I/O channel. When an atom position packet encounters the send-side cache, a lookup occurs to determine if the position has been previously stored in the cache. The upper portion of Figure 8 shows a case in which the position packet C[t] (representing the position of particle C from simulation time step t) arrives at the send-side cache and misses. A new entry is allocated if there is available space, but if no entry is available, the packet is not compressed. Even if the position is allocated, compression can not yet occur because the receive-side cache has no knowledge of the position, and thus, the entire position packet is transmitted over the I/O channel. The packet will also miss in the receive-side cache and is allocated the same entry.

On the subsequent time step (the lower portion of Figure 8), the position packet C[$t$+1] (i.e., the updated position for particle C) arrives at the send-side cache and hits. On a cache hit, the current particle position is compared to an estimate of the particle position based on the particle's position history. Because the receive-side cache has the same history and makes the same prediction, only the difference between the actual position and its estimated position needs to be sent across the channel. (Although we represent this as C[$t$+1] − C[$t$], that is a simplification of the actual prediction scheme used; details described in Section IV-B2). Static fields within the position packet are also available at the receive-side cache, so these can be replaced with a cache index. Ultimately, a special compressed position packet containing the cache index and the position difference (compressed using INZ) is transmitted over the I/O channel. On the arriving node, the INZ encoded position difference is decoded and the original copy of the position packet is reconstructed using information stored in the receive-side cache. It is important to note that the send-side cache and the receive-side cache always contain the exact same entries because they see the same cache access streams in the same order, and they are the same size and have the same organization.

The particle cache is four-way set associative with 1024 total entries. The eviction of entries from the particle cache is indirectly controlled by software. When a particle cache entry is hit by a position packet, it is marked with the current value of a time step counter maintained within the particle cache. This counter increments upon receipt of a special packet that the software sends to mark the end of the time step. Then, when a packet conflicts with a particle cache entry, that entry is evicted if the current time step counter value is greater than the entry's stored counter value by more than a specific (configurable) threshold. Otherwise, the particle cache is designed to be transparent to the software. The packet delivered to network endpoints will be the same regardless of whether that packet hit in any particle caches along its route.

*2) Position Extrapolation:* As described earlier, the particle cache takes advantage of the fact that particle positions in an MD simulation tend to follow smooth paths. By extrapolating a particle's future position and then only sending the difference between this extrapolation and the actual particle position, the number of bits sent over the I/O channels can be significantly reduced. The particle cache extrapolates each coordinate of a particle's position independently using a simple quadratic extrapolation.

Taking the $x$ coordinate of a particle's position as an example, let $x[t]$ be the particle's $x$ coordinate during time step $t$. The value of $x[t]$ can be accurately estimated (the estimate is denoted as $\hat{x}[t]$) with the second-order predictor using the particle's last position, velocity ($v$), and acceleration ($a$), which can be expressed using the coordinate's values from the previous three time steps, as follows:

$$v = \frac{x[t-1] - x[t-2]}{\Delta t},$$
$$a = \frac{(x[t-1] - x[t-2]) - (x[t-2] - x[t-3])}{\Delta t^2},$$
$$\hat{x}[t] = x[t-1] + \Delta t\,(v + a\,\Delta t)$$
$$= 3x[t-1] - 3x[t-2] + x[t-3].$$

Both the send-side cache and the receive-side cache store the particle's history, so the estimate $\hat{x}[t]$ is available to both. As a result, only the quantity $x[t] - \hat{x}[t]$ needs to be transmitted over the I/O channel. The same procedure is applied to the $y$ and $z$ coordinates of the particle's position.

The storage requirements for extrapolation can be reduced by first reformulating the estimator in terms of finite differences. To create our quadratic estimator, we store three differences—$D_0$, $D_1$, and $D_2$:

$$D_0[t] = x[t],$$
$$D_1[t] = x[t] - x[t-1],$$
$$D_2[t] = x[t] - 2x[t-1] + x[t-2].$$

Then, the estimate is simply the sum of these differences:

$$\hat{x}[t] = 3x[t-1] - 3x[t-2] + x[t-3]$$
$$= D_0[t-1] + D_1[t-1] + D_2[t-1].$$

Once an estimate is made, the differences are updated with the actual particle position according to:

$$D_0[t] = x[t],$$
$$D_1[t] = x[t] - D_0[t-1],$$
$$D_2[t] = x[t] - D_0[t-1] - D_1[t-1].$$

An advantage of this formulation is that the absolute values of $D_1[t]$ and $D_2[t]$ are expected to be small. Instead of 32 bits per coordinate, we store 12 bits of information per coordinate for the differences $D_1[t]$ and $D_2[t]$. When a new particle-cache entry is allocated, the estimator state is initialized using the current particle positions, and the differences $D_1[t]$ and $D_2[t]$ are simply set to zero. As more particle history is accumulated, the newly initialized estimator automatically transitions from a constant predictor to a linear predictor, and then to a quadratic predictor, without any special-case handling.

## C. Evaluation

INZ and the particle cache can be independently disabled in Anton 3, allowing us to measure the benefits of these compression schemes separately. For evaluation, we ran a synthetic water-only benchmark at various atom counts on a $2 \times 2 \times 2$ machine (i.e., 8 total nodes), with and without the compression features enabled. Figure 9a shows the decrease in transmitted bits over channels due to INZ alone, and due to both INZ and the particle cache combined. (Because the real hardware lacks performance counters for these statistics, the data in Figure 9a was collected from a detailed full-system simulator.) With only INZ enabled, the off-chip network traffic is reduced by 32%–40% across varying sizes of a water-only benchmark. While INZ alone is an effective and low-cost compression technique, using the particle cache in addition to INZ further reduces the number of transmitted bits at modest additional cost, ranging from 45% up to 62% of traffic reduction when compared to the baseline without any compression. The traffic reduction due to the particle cache decreases with larger atom counts because more atoms per node result in a higher cache miss rate. The size of the particle cache was chosen to provide sufficient traffic reduction for the low-atom-count regime, where the MD application is mostly communication-bound.

The increase in effective off-chip bandwidth provided by these compression techniques leads to improved application performance. Measured on real Anton 3 hardware, Figure 9b shows that the overall MD application speedup, with all compression enabled, ranges from 1.18 to 1.62.

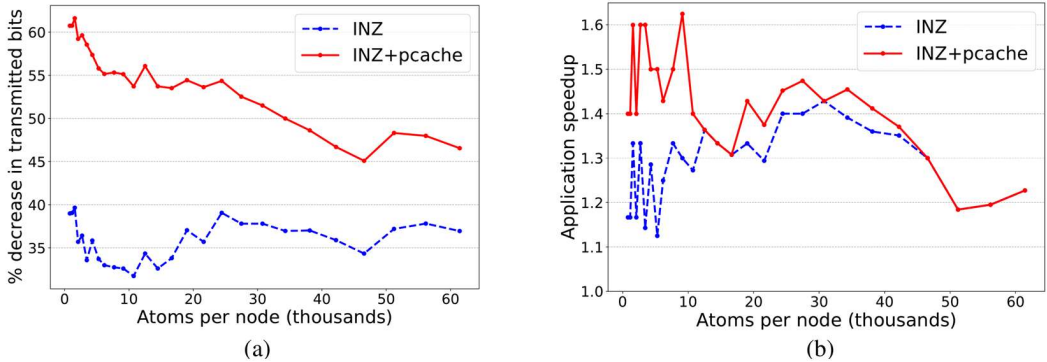

Figure 9. (a) Reduction in bits transmitted over channels due to INZ alone and INZ plus particle cache (pcache), measured using the architectural simulator; and (b) Overall application-level MD speedup measured under the same conditions on actual Anton 3 hardware.

## V. Fast Fine-grained Synchronization

The data flow for calculating pairwise interactions during Anton 3 simulations (described in Section II-C) requires synchronizing communication between a large number of source and destination pairs, with an unpredictable number of packets for each pair. Each ICB, for example, needs to be sure that it has received all the stream-set atoms (of unknown quantity, and from all GCs in the machine) before it notifies the PPIMs in its row that streaming is complete for a given time step; only then can the PPIMs start unloading their accumulated forces for stored-set atoms. Implementing this operation using a separate network packet between each pair would incur significant bandwidth cost that is proportional to the number of source and destination pairs.

To avoid this cost in Anton 3, we designed and implemented an in-network synchronization primitive, which we call a *network fence*. Network fences are implemented with *fence packets*; the receipt of a fence packet notifies the receiver that all packets sent before that fence packet have arrived. Fence packets are treated much like other packets, but with in-network merging and multicast support to reduce their bandwidth requirement. Each source component sends a fence packet after sending the packets it wants to arrive at destinations ahead of that fence packet. The network fence then guarantees that the destination components will receive that fence packet only after they receive all packets sent from all source components prior to that fence packet. The ordering guarantees for the network fence build upon the Anton 3 network's underlying ordering property (packets sent along a given path from source to destination are always delivered in the order in which they were sent), and the fact that a fence packet from a particular source is multicast along all possible paths a packet from that source could take to all possible destinations for that network fence.

### A. Software Interface

Anton 3 supports network fences for pre-defined pairs of source and destination component types (referred to as *fence patterns*), such as GC-to-ICB or GC-to-GC. Software running on a GC initiates a network fence with a two-argument instruction, `fence(pattern, number_of_hops)`, that specifies the fence pattern and the number of inter-node hops the fence will take through the torus network. For example, `fence(GC_to_ICB,3)` sends a 3-hop GC-to-ICB network fence. The receipt of this fence packet by an ICB indicates it has received all the position packets sent prior to this fence packet, from all GCs within three torus hops. This is a common use-case in MD simulation software because range-limited pairwise interactions only need to be computed between atoms that are within a fixed distance, and thus only require positions from remote ASICs within (at most) $k$ torus hops away from each node. By limiting the number of network hops, a network fence can achieve reduced latency for a limited synchronization domain (whereas setting `number_of_hops` to the network diameter will result in synchronization across the entire machine). It is important to note that all GCs must send fence packets to complete a given network fence; network fences do not support a subset of GCs in the machine participating in the fence operation.

### B. Fence Packet Merging and Multicast

Without in-network merging of fence packets, implementation of the logical concept of network fence would cause significant bandwidth cost. In the Anton 3 network, fence packets are thus combined by the network routers. Below, we describe fence merging and multicast mechanisms for a single network fence; support for multiple concurrent fences will be discussed in Section V-D.

Figure 10a illustrates how fence merging is achieved at router input ports. When a fence packet arrives, instead of forwarding the packet to the output port, the input port merges the fence packet. This is implemented by incrementing a *fence counter*; when the fence counter reaches the expected value, a single fence packet is transmitted to each output port. A *fence output mask* determines the set of output ports that the fence should be multicast to; for input port $i$, bit $j$ of its output mask is set if the fence packet needs to travel from the input port $i$ to the output port $j$ within that router. When the fence packet is sent out, the counter is reset to zero. Because the router can continue forwarding non-fence packets while it is waiting for the last arriving fence packet, normal traffic sent after the fence packet may reach the destination before the fence packet (i.e., the network fence works as a one-way barrier).

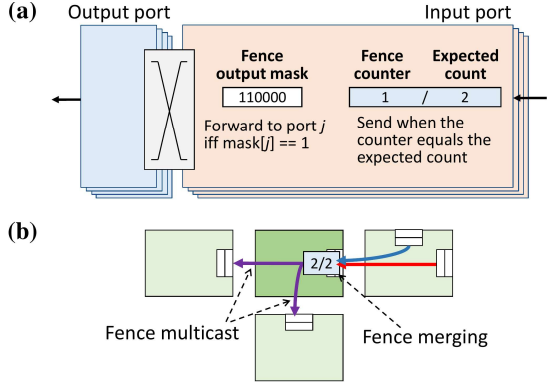

Figure 10. (a) Fence merging and multicast within a single network router. (b) An example of the routes that fence packets take through multiple routers.

The expected count and the fence output mask are preconfigured by software for each fence pattern. For the example in Figure 10b, the particular input port of the middle router expects fence packets from two different paths in the upstream router. Because one fence packet will arrive from each path due to merging, the input port will receive a total of two fence packets, thereby setting the expected count to two. The fence counter width is limited by the number of router ports (e.g., $\lceil \log_2(6+1) \rceil = 3$ bits for a six-port router). The fence output mask in this example will have two bits set for the two output ports to which the fence packets are multicast.

### C. Implementing a Network Fence through the Inter-Node Torus Network

The routing algorithm for the inter-node torus network exploits the path diversity from six possible dimension orders, as well as two physical channel slices for each connected neighbor. In addition, multiple VCs are employed to avoid network deadlock in the 3D torus, meaning that fence packets must be sent to all possible VCs along the valid routes that packets can travel. When the network fence crosses the channel, fence packets are thus injected to the Edge Network by the Channel Adapter on all possible request-class VCs. Although some hops may not necessarily utilize all of these VCs, this rule ensures that the network fence covers all possible paths throughout the entire network and simplifies the fence implementation because an identical set of VCs can be used regardless of the number of hops the packet has taken. Within the Edge Router, a separate fence counter must be used for each VC; only the fence packets from the same VC can be merged.

### D. Concurrent Network Fences

Up to this point, we have only described the implementation of a single network fence in the network. By adding more fence counters in routers, the Anton 3 network supports concurrent outstanding network fences, allowing software to overlap multiple fence operations (up to 14). To reduce the size requirement for the fence counter arrays in the Edge Router, the network adapters (the Channel Adapters and the Row Adapters) implement flow-control mechanisms, which control the number of concurrent network fences in the Edge Network by limiting the injection of new network fences. These flow-control mechanisms allow the network fence in Anton 3 to be implemented using only 96 fence counters per input port of the Edge Router.

### E. Global Barrier

A network fence with a GC-to-GC pattern can be used as a barrier to synchronize all GCs within a given number of torus hops; once a GC has received a fence, then it knows that all other GCs have sent one. When the number of inter-node hops for a GC-to-GC network fence is set to the machine diameter, it behaves as a global barrier.

When a GC-to-GC network fence arrives at the GC, it translates into a counted write (described in Section III-A) to a specified memory address within the local SRAM with a fixed count. After sending a GC-to-GC fence, each GC can thus issue a blocking read to the memory location with the same counter threshold to detect the arrival of the fence packet, and the blocking read acts as a synchronization barrier point (as the blocked read will only be unblocked after fences from all GCs have arrived). It is important to note that the synchronization barrier implemented with a GC-to-GC network fence also guarantees that all writes to remote SRAMs from GCs are complete, as the network fence travels all the valid paths for those writes.

### F. Evaluation

To evaluate the performance of network fence, we measured the barrier synchronization latency on a real 128-node Anton 3 machine using a GC-to-GC network fence across varying hop counts of the network fence (Figure 11). The 0-hop case represents the intra-node barrier where the network fence does not travel over off-chip channels, and this takes about 51.5 ns. The 8-hop case represents the global barrier across the entire 128-node Anton 3 machine (connected as a 4 × 4 × 8 torus), and this takes only about 504 ns. With multiple nodes, the latency comprises approximately 91.2 ns of fixed overhead and 51.8 ns of per-hop latency. This per-hop latency is about 17.6 ns longer than the average per-hop latency of inter-node communication reported in Section III-C, because fence packets need to travel through all the valid paths at every hop. Overall, the results show that the Anton 3 network provides a high-performance synchronization mechanism that can synchronize a large number of source-destination pairs with a latency close to the one-way messaging latency between the pairs. The network fence barrier latency also scales linearly with the network diameter.

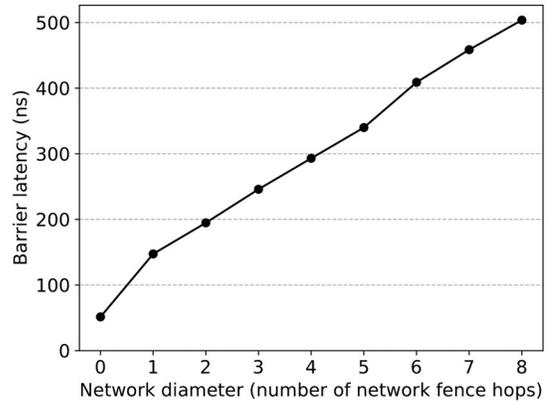

Figure 11. Network fence barrier latency, measured on a real 128-node Anton 3 machine.

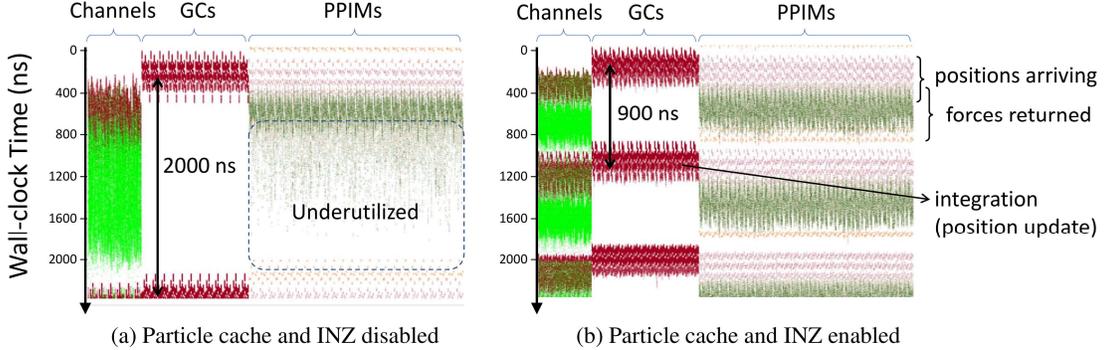

Figure 12. Machine activity plots (data collected from the architectural simulator) showing activity across the ASIC for range-limited pairwise interactions during simulation of a water-only system on an 8-node Anton 3, with compression schemes either disabled (a) or enabled (b). A time step takes roughly 2000 ns with compression disabled, and 900 ns with compression enabled. Columns representing inter-node network traffic over the channels are shown on the left of each plot (with position packets in red and force packets in green), the middle columns show integration activities across GCs, and the rightmost columns show position packets arriving at PPIMs and force packets being returned.

## VI. Performance and Cost Analyses

### A. Machine Activity during MD Simulation

Figure 12 plots a portion of machine activity while computing range-limited pairwise interactions (for roughly 2500 ns of wall-clock time) from a 32,751-atom, water-only benchmark on an 8-node Anton 3 machine (data collected from a detailed full-system simulator). Panel (a) shows the data with compression features (INZ and particle cache) disabled, and panel (b) shows the data with those features enabled. Each column represents a different hardware component, and each color represents a different type of computation or network traffic.

As shown in Figure 12a, the inter-node channels are heavily utilized while the primary compute resources for pairwise interactions (PPIMs) remain underutilized, thus motivating our network specializations for bandwidth reduction. Through our novel compression schemes, the amount of time required to send packets over the network can be significantly reduced (as shown in Figure 12b), and the compression also leads to more efficient utilization of the PPIMs. As a result, each simulation time step takes about 900 ns with compression enabled, as opposed to roughly 2000 ns with compression disabled. In addition, the fact that each time step takes only around 900 ns (~2500 cycles) with compression enabled indicates that any synchronization must have very low overhead. In Anton 3, we address this need by implementing the network fence and fine-grained synchronization with counted write and blocking read. These network features contribute to Anton 3's overall MD performance improvement, which is substantial; simulation of a 2.2-million-atom ribosome, for example, is roughly 19 times faster on a 512-node Anton 3 than on a 512-node Anton 2 (while consuming approximately one-tenth the energy for a given simulation), and is 460 times faster than on any general-purpose machine (comprehensive performance comparisons available in [6]).

### B. Component Area

The Anton 3 network comprises four distinct component types (excluding SERDES IPs and IP-specific support logic): (1) the Core Router, (2) the Edge Router, (3) the Channel Adapter, and (4) the Row Adapter. Table II shows the individual contributions of these network components to the total die area, indicated as a percentage of the floorplan area; about 14.1% of the ASIC's total area is used by the network components.

TABLE II: Network component contributions to the total die area.

| Network component | Component count | % of total die area |
|---|---|---|
| Core Routers | 288 | 9.4% |
| Edge Routers | 72 | 1.4% |
| Channel Adapters | 24 | 2.8% |
| Row Adapters | 72 | 0.5% |
| Total | | 14.1% |

We also examined the implementation costs of the particle cache and network fence. The major cost of the particle cache stems from on-die memory for cache storage in each Channel Adapter, and the major cost of the network fence arises from fence counter arrays included in all the network routers. Table III lists the costs of these network features in terms of die area. The total implementation costs for both the particle cache and network fence amount to only 1.8% of the total die area—a small overhead considering the performance benefits from these features.

TABLE III: IMPLEMENTATION COSTS OF NETWORK FEATURES IN THE ANTON 3 ASIC.

| Network feature | % of total die area |
|---|---|
| Particle Cache | 1.6% |
| Network Fence | 0.2% |
| **Total** | **1.8%** |

## VII. RELATED WORK

*MD Simulation*: Over the past few decades, use of MD simulation has become increasingly popular in the fields of molecular biology and drug discovery [19]–[22]. In order to achieve simulation timescales long enough to capture many interesting biochemical processes, research has focused on accelerating simulations by improving their underlying algorithms [23]–[25] and parallelizing them in commodity hardware like GPUs [26]–[28] or FPGAs [29][30]. Some researchers have scaled MD software to run on general-purpose supercomputers [31]–[33], while others have built supercomputers specialized for MD (such as the MDGRAPE series [34]–[36] and the Anton series [3]–[6]).

*Compression*: Many data compression techniques have been proposed to increase effective cache size and memory bandwidth. The compression cache [37], for example, replaces frequently accessed data in a cache with indices to a table containing the values. Another technique, significance-based compression [38][39], focuses on the data that contain information in a few low-order bits, and thus encodes data to fewer bits. Base-Delta-Immediate compression [40] exploits the low dynamic range of values in cache lines, and compresses them using a base and deltas. Bit-Plane Compression [41] first transforms data to improve the compressibility of data by increasing the run-length of zeros, similar to INZ in Anton 3. Like these earlier approaches, the compression schemes in Anton 3 make use of data characteristics, but are implemented specifically to reduce off-chip traffic in MD simulations. Such domain-specific approaches are also common in GPUs [42][43] and machine learning (ML) accelerators [44][45].

*Synchronization*: Several large-scale parallel machines provide specialized hardware support to accelerate collective operations, such as synchronization barriers. Blue Gene/L [46], Blue Gene/P [47], and the Cray T3D [48] contain a dedicated network for global barriers. Blue Gene/Q [49] and the Cray T3E [50], on the other hand, embed a virtual tree network into the regular network, and use special packets and in-network logic to implement barriers. Using a special fence packet and extra logic in routers for merging and multicast, the Anton 3 network implements an *all-to-all* barrier on the regular network that achieves low latency and scales linearly with respect to the network diameter. A similar all-to-all barrier on an on-chip mesh network was proposed previously [51]; in that approach, however, packets that are simultaneously in-flight within a given router are merged opportunistically, and only synchronization within a given chip is addressed. One important difference between the synchronization barrier in Anton 3 and those described in the works listed above is that the barrier in Anton 3 is supported by the network fence, which was originally designed to enforce packet ordering. As a result, unlike the previously proposed schemes, the barrier in the Anton 3 network also works as a memory fence, thus guaranteeing that all writes from GCs to remote on-chip memories are complete across the entire machine.

## VIII. CONCLUSION

As parallel MD simulation requires frequent inter-node communication, simulation performance can be limited by both off-chip bandwidth and latency. Due to the slow scaling of off-chip bandwidth in low-latency networks in current semiconductor technology, a high-performance network is critical to maximize performance in large-scale parallel systems. In Anton 3, we have designed and implemented a tightly integrated network that provides fast end-to-end inter-node communication and synchronization for fine-grained messages. As a result of various design choices in network components to minimize network latency, and implementing blocking read synchronization to reduce the arrival-to-use latency for data received over the network, the end-to-end one-way latency between cores can be as low as 55 ns for neighboring nodes.

The Anton 3 network also increases effective off-chip bandwidth by using two MD-specific compression techniques: INZ (which efficiently compresses payloads with small absolute values) and the particle cache (which allows chips to transmit position differences over the off-chip channels instead of entire positions). Lastly, the network implements a novel synchronization mechanism called a network fence, which supports low-latency, fine-grained synchronization across a large number of compute units. Through in-network merging and multicast support, the bandwidth requirement of the network fence is significantly reduced. These features, implemented with very little chip area, improve the utilization of the on-chip compute resources, helping Anton 3 to massively speed up MD simulations across a range of problem and machine sizes. We also believe that, although these network features were designed specifically for MD simulation, the underlying concepts are applicable to many other high-performance parallel applications.

## ACKNOWLEDGMENTS

We thank Larry Nociolo for his work on signal integrity analysis for high speed channels of Anton 3; Jeffrey S. Kuskin for implementing memory modules with blocking read support; Alistair Bell, Andrew Parker, Michael Wazlowski, and Michael Theobald for their work on the verification of the Anton 3 network components; Ken Mackenzie for assistance in running experiments; Michael Fenn for valuable discussions; Adam Butts and Mark Moraes for a critical reading of the manuscript; and Eric Martens and Berkman Frank for editorial assistance.